\documentclass[11pt,letterpaper]{article}
\usepackage[utf8]{inputenc}
\usepackage{amsmath, amsfonts, amssymb}
\usepackage{verbatim}
\usepackage{graphicx}
\usepackage{mathrsfs}
\usepackage{cite}
\usepackage{physics}
\usepackage{hyperref}
\usepackage[margin=2.5cm]{geometry}
\usepackage{multicol}
\usepackage{setspace}
\usepackage{secdot}
\usepackage{caption}
\usepackage{subcaption}
\usepackage{indentfirst}
\usepackage{xcolor}

\newcommand{\ho}{h_0}
\newcommand{\hl}{h_1}
\newcommand{\dA}{A_v}
\newcommand{\vq}{q}

\newcommand{\cT}{\mathcal{T}}

\newcommand{\ca}{a}
\newcommand{\cb}{b}

\newcommand{\New }[1]{{\color{black} #1}}
\newcommand{\NewY}[1]{{\color{black} #1}}
\newcommand{\NewM}[1]{{\color{black} #1}}

\newcommand{\refcom}[1]{{\color{black} #1}} 

\title{\sffamily\bfseries Odd-parity perturbations in the most general scalar-vector-tensor theory}
\author{{\it Yolbeiker Rodr\'{\i}guez Baez
	\footnote{Email: \href{mailto:yolbeiker.rodriguez@pucv.cl}{yolbeiker.rodriguez@pucv.cl}}, 
	Manuel Gonzalez-Espinoza.\footnote{Email: \href{mailto:manuel.gonzalez@pucv.cl}{manuel.gonzalez@pucv.cl}}}\\
\\
Instituto de F\'{\i}sica, Pontificia Universidad Cat\'olica de Valpara\'{\i}so,\\ 
Av. Brasil 2950, Valpara\'{\i}so, Chile\\}
\date{April, 2023}

\begin{document} 
\maketitle
\flushbottom

\begin{abstract}
In the context of the most general scalar-vector-tensor theory, we study the stability of static spherically symmetric black holes under linear odd-parity perturbations. 
We calculate the action to second order in the linear perturbations to derive a master equation for these perturbations. 
For this general class of models, we obtain the conditions of no-ghost and Laplacian instability. 
Then, we study in detail the generalized Regge-Wheeler potential of particular cases to find their stability conditions.
\end{abstract}

\section{Introduction}

The main reasons to modify General Relativity (GR), one of the most successful theories in physics, come from direct cosmological observations and strong theoretical arguments. For instance, on the large-scale regime, GR cannot clearly explain (without introducing other issues) the late-time accelerated expansion of the Universe \cite{SupernovaSearchTeam:1998fmf, SupernovaCosmologyProject:1998vns}. Conversely, GR needs to be modified in the high-energy limit to solve singularities inside black holes \cite{Hawking:1970zqf}. Since all attempts to quantize gravity and to unify it with the other interactions of nature have failed, modification of the fundamental theories is somewhat inevitable. Options to modify GR range from introducing new fundamental fields, abandoning the diffeomorphism invariance, violating the weak equivalence principle, and considering extra dimensions (see \cite{Capozziello:2011et, Berti:2015itd} for a review).
 
Adding extra degrees of freedom (DOFs) to Einstein-Hilbert action seems to be the most promising way to modify GR \cite{Clifton:2011jh}. However, introducing new fundamental fields with a nontrivial coupling to gravity is not an easy task. Any modification of a theory must respect basic principles to describe physical situations. We demand that coupling of new DOFs does not produce Ostrogradsky instability \cite{Ostrogradsky:1850fid, Woodard:2015zca}, and theory must have a well-posed Cauchy problem.
Furthermore, it is well-known that when adding a scalar field to the Einstein-Hilbert action, Hordenski's theories \cite{Horndeski:1974wa} are the most general scalar-tensor theories with second-order equations of motion. Nevertheless, if instead of a single scalar degree of freedom, we add a vector field that couples to gravity, generalized Proca theories \cite{Tasinato:2014eka, Heisenberg:2014rta, Allys:2015sht} are the most general vector-tensor theories with second-order equations of motion.
However, we might add more than one fundamental field at once. 
In this context, scalar-tensor and vector-tensor theories can be unified in scalar-vector-tensor (SVT) theories. 
The most general action that couples a scalar ($\phi$) and a vector field ($A_\mu$) with gravity was constructed carefully in Ref. \cite{Heisenberg:2018acv} respecting locality, unitarity, Lorentz invariance, and keeping a pseudo-Riemannian geometry. 
As shown in the original paper, these theories can be classified into two groups depending on the symmetries of the vector field. 
When the vector field respects the $U(1)$ gauge invariance, we have (in total) five DOFs: the usual two tensor polarizations, the additional two transverse vectors, and one scalar DOF. Alternatively, when the $U(1)$ gauge symmetry is broken, it leads to the propagation of a longitudinal scalar besides the five DOFs.

After the proposal of SVT theories, their direct implications in cosmology and astrophysics have recently been investigated. 
For example, using a Friedmann-Lemaitre-Robertson-Walker (FLRW) background, Ref. \cite{Heisenberg:2018mxx} studied conditions for the absence of ghosts and Laplacian instability associated with tensor, vector, and scalar perturbations at linear order.
This research has been extended in Ref. \cite{Kase:2018nwt} by adding a perfect fluid in order to study applications to dark energy\footnote{SVT theories have been widely used as alternatives to dark matter particles. Predominant examples are TeVeS \cite{Bekenstein:2004ne} and MOG \cite{Moffat:2005si}.}.
Also, SVT theories have been studied in inflation (see, for instance, \cite{Oliveros:2022njz}).

Since the detection of gravitational waves by the LIGO-Virgo collaboration \cite{LIGOScientific:2016aoc, LIGOScientific:2016sjg, LIGOScientific:2017bnn, LIGOScientific:2017vox, LIGOScientific:2020iuh, LIGOScientific:2021qlt}, one of the most promising ways to detect new DOFs in nature and, therefore, to detect a violation of the non-hair theorem, is by analyzing the quasinormal modes of Black Holes (BH), (see \cite{Isi:2019aib, Bhagwat:2019dtm, LIGOScientific:2020tif} and references therein for works that constraint GR solutions with gravitational-wave data from the LIGO detections).
Through linear perturbation theory, apart from getting the quasinormal mode frequency, we can study the stability of new BH solutions and conditions for the absence of ghosts and Laplacian instabilities when new DOFs are added to the action.
This formalism, originally developed by Regge-Wheeler \cite{Regge:1957td} and Zerilli \cite{Zerilli:1970se} (see also \cite{Chandrasekhar98}), which demonstrated the stability of the Schwarzschild black hole, has been applied to many other theories and BH solutions, such as, BHs in self-gravitating nonlinear electrodynamics \cite{Moreno:2002gg}, Hordenski theories \cite{Kobayashi:2012kh, Kobayashi:2014wsa, Ganguly:2017ort}, $f(R,G)$ gravity models \cite{DeFelice:2011ka}, BHs in generalized Einstein-Maxwell-scalar theories \cite{Gannouji:2021oqz}, generalized Proca theories \cite{Chagoya:2016aar, Heisenberg:2017hwb, Kase:2018voo}.
For SVT theories, this formalism has been applied in Ref \cite{Heisenberg:2018mgr}. They studied properties of BHs on a static and spherically symmetric background with $U(1)$ gauge symmetry, i.e., a subgroup of the general action.

The current paper extends the above results to study BH perturbations for the full set of SVT theories.
\refcom{Determining the stability and getting the quasinormal modes frequencies of BH given by this theory is the first step to constraining the Lagrangian with future detections of gravitational waves.}
In this paper, we focus on the stability of BHs under odd-parity perturbations since the analysis of even-parity modes is generally more laborious. We leave the later analysis to future research.
\refcom{As we will state later, this analysis is performed in a spherically symmetric spacetime. Specifically, we focused on BH solutions where the non-zero component of the vector field is associated with the electric charge.}

This paper is organized as follows. In section \ref{section:SVT_review}, we first review the most general action for SVT theories. Under a spherically symmetric background, in section \ref{section:BG_equations}, we derive the background equations of motion. In section \ref{section:review_pertubation}, we summarize the formalism of black hole perturbation theory focusing on odd-parity modes. We derived the second-order Lagrangian for the dipole and higher multipoles. With this analysis, we get conditions for the absence of ghosts and Laplacian instabilities and derive the propagation speed of the available DOFs. Finally, we will apply our results to interesting examples, comparing them with previous results found in the literature, before final comments and conclusions.

To facilitate the use of our results, a {\scshape Mathematica}\textsuperscript{\textregistered} notebook is available online \cite{mathematica}.

\section{Review of SVT theories \label{section:SVT_review}}

In this paper, we are interested in studying BH perturbations in the most general SVT theory \NewY{having second-order 
equations} \NewM{of motion.}
As shown in \cite{Heisenberg:2018acv}, \NewY{the most general action, with broken $U (1)$ gauge invariance, is the following}
\begin{equation}
    S_{\rm SVT}= \int \dd[4]{x} \sqrt{-g}\, \sum_{n=2}^6 \mathcal{L}_{{\rm SVT}}^{(n)}\,,
    \label{action}
\end{equation}
with Lagrangians
\begin{equation}
\begin{aligned}
    \mathcal{L}_{{\rm SVT}}^{(2)} & = 
        f_2(\phi,X_1,X_2,X_3,F,Y_1,Y_2,Y_3) \,,\\
    \mathcal{L}_{{\rm SVT}}^{(3)} & = 
        f_{3}(\phi,X_3)g^{\mu\nu}S_{\mu\nu} + \tilde{f}_{3}(\phi,X_3)A^{\mu}A^{\nu} S_{\mu\nu} \,,\\
    \mathcal{L}_{{\rm SVT}}^{(4)} & = 
        f_{4}(\phi,X_3)R+f_{4,X_3}(\phi,X_3) \left[(\nabla_\mu A^\mu)^2-\nabla_\mu A_\nu \nabla^\nu A^\mu \right] \,,\\
    \mathcal{L}_{{\rm SVT}}^{(5)} & = 
        f_5(\phi,X_3)G^{\mu\nu} \nabla_{\mu}A_{\nu} 
        - \frac{1}{6}f_{5,X_3}(\phi,X_3) \left[ (\nabla_{\mu} A^{\mu})^3 - 3\nabla_{\mu} A^{\mu} 
        \nabla_{\rho}A_{\sigma} \nabla^{\sigma}A^{\rho} 
        \right. \\ &  \left. 
        + 2\nabla_{\rho}A_{\sigma} \nabla^{\gamma} A^{\rho} \nabla^{\sigma}A_{\gamma}\right] 
        + \mathcal{M}_5^{\mu\nu}\nabla_\mu \nabla_\nu\phi
        + \mathcal{N}_5^{\mu\nu}S_{\mu\nu} \,,\\
    \mathcal{L}_{{\rm SVT}}^{(6)} & = 
        f_6(\phi,X_1)L^{\mu\nu\alpha\beta}F_{\mu\nu}F_{\alpha\beta}
        + \mathcal{M}_6^{\mu\nu\alpha\beta}\nabla_\mu\nabla_\alpha \phi\nabla_\nu\nabla_\beta\phi
        + \tilde{f}_6(\phi,X_3)L^{\mu\nu\alpha\beta}F_{\mu\nu}F_{\alpha\beta} 
        \\ &
        + \mathcal{N}_6^{\mu\nu\alpha\beta}S_{\mu\alpha}S_{\nu\beta} \,,
	\end{aligned}
	\label{eq:Lag_SVT}
\end{equation}
where $g$ is the determinant of the metric tensor $g_{\mu\nu}$, $R$ is the Ricci scalar, $G^{\mu \nu}$ is the Einstein tensor, and $\nabla_\mu$ is the covariant derivative operator. In the above Lagrangians, the scalar quantities 
\begin{equation}
    X_1=-\frac{1}{2} \nabla_{\mu} \phi \nabla^{\mu} \phi\,,\qquad
    X_2=-\frac{1}{2} A^{\mu} \nabla_{\mu} \phi \,,\qquad 
    X_3=-\frac{1}{2} A_{\mu} A^{\mu}\,,
    \label{X123}
\end{equation}
represent the kinetic term of the scalar field, an interaction between the fields, and the mass term for the vector field. 
\refcom{The new terms $X_2$ and $X_3$ and the functions with these dependencies, which are not present in \cite{Heisenberg:2018mgr}, clearly break the $U(1)-$symmetry of the theory.}
\NewY{Also, for the vector field,} we have considered its standard kinetic term and introduced the antisymmetric field strength tensor $F_{\mu \nu}$,
and its dual $\tilde{F}_{\mu \nu}$, defined as 
\begin{equation}
    F=-\frac{1}{4}F_{\mu\nu}F^{\mu\nu}\,,\qquad
    F_{\mu\nu}=\nabla_\mu A_\nu-\nabla_\nu A_\mu\,,     \qquad 
    \tilde{F}^{\mu\nu}=\frac{1}{2}
    \mathcal{E}^{\mu\nu\alpha\beta}F_{\alpha\beta}\, ,
\end{equation}
and several Lorentz-invariant quantities associated with intrinsic vector modes, as
\begin{equation}
	Y_1 = \nabla_\mu \phi \nabla_\nu \phi\,F^{\mu\alpha}F^\nu{}_\alpha \,, \qquad 
	Y_2 = \nabla_\mu\phi\, A_\nu F^{\mu\alpha}F^\nu{}_\alpha \,, \qquad
	Y_3 = A_\mu A_\nu F^{\mu\alpha}F^\nu{}_\alpha\,.
	\label{FY123}
\end{equation}

\refcom{When removing the restriction of $U(1)$-gauge invariance, the vector field also enters in the action through its symmetric part 
\begin{equation}
	S_{\mu \nu}=\nabla_{\mu}A_{\nu} + \nabla_{\nu}A_{\mu}
\end{equation}
and, therefore, permits the construction of more general interactions. To find the exact interactions of this new action it is convenient to introduce effective metrics related to each interaction, $\mathcal{G}_{\mu\nu}^{k}$, which depend on possible combinations of $g_{\mu\nu}$, $\nabla_\mu \phi$ and $A_\mu$:
\begin{equation}
   \mathcal{G}_{\rho\sigma}^{k} = 
	 k_{1}(\phi,X_i)g_{\rho\sigma}+k_{2}(\phi,X_i) \nabla_\rho \phi \nabla_\sigma \phi
   + k_{3}(\phi,X_i)A_\rho A_\sigma
   + k_{4}(\phi,X_i)A_\rho \nabla_\sigma \phi.
   \label{eq:effctive_metric}
\end{equation}
For example, for the cubic interactions, the general form of the effective metric must be restricted to $\mathcal{G}_{\mu\nu}^{f_3} =f_{3}(\phi,X_3)g^{\mu\nu} + \tilde{f}_{3}(\phi,X_3)A^{\mu}A^{\nu}$ to obtain second order equations of motion, for more details see \cite{Heisenberg:2018vsk}. On the other hand, for the fifth-order Lagrangian case, we need two general forms of the effective metric and a non-minimal coupling to the Einstein tensor to obtain the most general scalar-vector-tensor theory without Ostrogradsky instability \cite{Heisenberg:2018vsk}. 
Accordingly, the effective tensor $\mathcal{G}_{\mu\nu}^{k}$ enters into the Lagrangian (\ref{eq:Lag_SVT}) through the definition of the rank-2 tensors $\mathcal{M}^{\mu\nu}_5$ and $\mathcal{N}^{\mu\nu}_5$ in ${\cal L}_{\rm SVT}^{(5)}$, which are associated with intrinsic vector modes, and are defined as
\begin{equation}
    \mathcal{M}^{\mu\nu}_5
        = \mathcal{G}_{\rho\sigma}^{k} \tilde{F}^{\mu\rho}\tilde{F}^{\nu\sigma}\,,
    \qquad\quad 
    \mathcal{N}^{\mu\nu}_5
        = \mathcal{G}_{\rho\sigma}^{\tilde{k}} \tilde{F}^{\mu\rho}\tilde{F}^{\nu\sigma}\,,
\end{equation}
where $\mathcal{G}_{\rho\sigma}^{\tilde{k}}$ has the same definition as (\ref{eq:effctive_metric}) with a tilde in the functions $k_i$.
In general, the functions $k_{j}$ and $\tilde{k}_{j}$ (where $j=1,2,3,4$) depend on $\phi$ and $X_1, X_2, X_3$. 
However, as shown in \cite{Heisenberg:2018acv, Kase:2018nwt}, to maintain the non-dynamic nature of the temporal component of $A_\mu$ in arbitrary curved backgrounds, it is essential that the dependence of either $X_1$ or $X_3$ in $k_{j}$ and $\tilde{k}_{j}$ is manifestly dominant.}

In the Lagrangian (\ref{eq:Lag_SVT}), we also introduced the double dual Riemann tensor defined by
\begin{equation}
    L^{\mu\nu\alpha\beta} = \frac{1}{4} \mathcal{E}^{\mu\nu\rho\sigma} \mathcal{E}^{\alpha\beta\gamma\delta} R_{\rho\sigma\gamma\delta}\,,
\end{equation}
where $\mathcal{E}^{\mu\nu\alpha\beta}$ is the completely anti-symmetric Levi-Civita tensor 
obeying the normalization $\mathcal{E}^{\mu\nu\alpha\beta}
\mathcal{E}_{\mu\nu\alpha\beta}=-4!$, and $R_{\rho\sigma\gamma\delta}$ is the Riemann tensor. 
Finally, the rank-4 tensors $\mathcal{M}_6^{\mu\nu\alpha\beta}$ and $\mathcal{N}^{\mu\nu\alpha\beta}_6$,
in the sixth order Lagrangian, are defined, respectively, by 
\begin{equation}
    \mathcal{M}^{\mu\nu\alpha\beta}_6 =2f_{6,X_1} (\phi, X_1) \tilde{F}^{\mu\nu}\tilde{F}^{\alpha\beta} \,, \qquad
    \mathcal{N}^{\mu\nu\alpha\beta}_6 =\frac12\tilde{f}_{6,X_3} (\phi, X_3) \tilde{F}^{\mu\nu}\tilde{F}^{\alpha\beta}\,.
\end{equation}

As common in the literature, we have denoted partial derivatives of the Lagrangian functions $f_i$ with respect to 
$x = \{\phi,\, X_i,\, F, \, Y_i\}$ as $f_{i, x} \equiv \partial f_i/\partial x$.

\section{Background equations \label{section:BG_equations}}

We assume a static and spherically symmetric background given by the metric\footnote{We will use the notation $\bar{X}$ to refer to quantities evaluated at the background level.}
\begin{equation}
    d s^2 = \bar{g}_{\mu\nu} \dd x^\mu \dd x^\nu = -A(r) \dd t^2 + B^{-1}(r) \dd r^2 + C(r)^2( \dd \theta^2 + \sin^2{\theta} \dd \varphi^2)
    \label{eq:BG-metric}
\end{equation}
where $A(r)$, $B(r)$, and $C(r)$ are functions of the radial coordinate $r$.
To satisfy the background symmetries of our spacetime, we will consider that the scalar field only depends on the radial coordinate, 
$\bar{\phi} \equiv \phi(r)$, and the vector field takes the form
\begin{equation}
    \bar{A}_\mu = (A_0(r), A_1(r), 0,0)\, .
\end{equation}
where $A_0$ and $A_1$ depends only on the $r$ coordinate.
\NewY{As stated in \cite{Gannouji:2021oqz}, this choice of the vector field is not the most general form in a spherically symmetric spacetime. 
With this choice, we might only study electric charge since the contribution of the magnetic charge has been neglected.
}
In this background, the interactions terms are given by
\begin{equation}
    \bar{X}_1 = -\frac{1}{2} B \phi '^2   					\, , \qquad
    \bar{X}_2 = -\frac{1}{2} A_1 B \phi ' 					\, , \qquad
    \bar{X}_3 = \frac12 \qty(\frac{{A_0}^2}{A}- A_1^2 B) 	\, , \qquad
    \bar{F}   = \frac{B A_0'^2}{2 A} \, ,
\end{equation}
where a prime represents a derivative with respect to $r$.

Replacing the metric (\ref{eq:BG-metric}), the scalar and vector fields in the action (\ref{action}), we obtain
the equations of motion (EOM) $\mathcal{E}_X = 0$, by varying the action with respect to the functions 
$X = \{A,\, B,\, C,\, \phi,\, A_0,\, A_1\}$.
\NewY{The explicit form of the EOMs are long expressions and do not provide additional information. 
Therefore, we omit the expressions here and provide them in a companion Mathematica notebook \cite{mathematica}.
As a further remark, the EOM associated with the vector component $A_0$ and  the scalar field can be written as 
\begin{equation}
    \mathcal{E}_{A_0} = -\dv{r}(\sqrt{\frac{B}{A}} \mathcal{J}_{A_0}) - \sqrt{\frac{B}{A^3}} \mathcal{S}_{A_0}
    \,, \qquad
    \mathcal{E}_{\phi} = \dv{r}(\sqrt{\frac{B}{A}} \mathcal{J}_\phi) + \sqrt{\frac{B}{A}} \mathcal{S}_\phi
    \, ,
\end{equation}
from which, and for some especial theories, we might derive the conserved charges.}
As we pointed out, the function $f_2$ (in the second order Lagrangian (\ref{eq:Lag_SVT})) depends on $\{\phi, \, X_i,\, F,\, Y_i\}$, where $i=1,2,3$. 
However, on the static and spherically symmetric background (\ref{eq:BG-metric}) we have $Y_i = 4 X_i F$, so it is redundant to \NewY{include this last dependency} on $f_2$. \NewY{In consequence, from now on,} we will consider that $f_2 \equiv f_2(\phi, X_1, X_2, X_3, F)$.

\section{Perturbation formalism: odd-parity modes \label{section:review_pertubation}}

Given the background equations of motion, we proceed to derive the quadratic action for perturbations. 
As is well known in the literature, due to the pioneering work of Regge-Wheeler \cite{Regge:1957td} and later extended by Zerilli \cite{Zerilli:1970se}, metric perturbations around a spherically symmetric 4-dimensional spacetime can be decomposed into odd-parity and even-parity modes according to their transformation properties under a two-dimensional rotation of the sphere. 
Even though the Regge-Wheeler formalism has been initially applied to the Schwarzschild spacetime, the formalism relies on the spacetime symmetries, so, with the metric (\ref{eq:BG-metric}), it can be applied to general theories, including SVT theories.

Since the action (\ref{action}) does not contain a Chern-Simons term or a Pontryagin density (in our Lagrangian the totally antisymmetric tensor $\mathcal{E}^{\alpha\beta\gamma\delta}$ always is present with even powers),
the SVT action (\ref{action}) does not violate the parity transformation of perturbations
(see \cite{Yunes:2007ss, Motohashi:2011pw} for more details on BH perturbations in parity violating gravitational theories).
Thus, at the level of the action odd-parity and even-parity modes decouple from each other, allowing one to study each mode separately.

In this paper, we will study the odd-parity perturbations\footnote{These type of perturbations are also known in the literature as vector or axial perturbations, since under parity transformation $(\theta, \varphi) \to (\pi - \theta, \pi + \varphi)$ these modes pick a factor $(-1)^{\ell +1}$.}. 
\NewM{And, due the fact that a scalar field does not contribute to odd-parity perturbations, we will consider linear perturbations only for the metric and the vector field}
\begin{equation}
    g_{\mu \nu} = \bar{g}_{\mu \nu} + h_{\mu \nu} \, , \qquad\qquad A_\mu = \bar{A}_\mu + \delta A_\mu \, .
\end{equation}
It can be checked that only 3 of the 10 components of the perturbed metric correspond to odd-parity modes. In fact, under a reparametrization of the angles \NewM{$\theta$ and $\varphi$} into $R_2(\theta,\varphi)$ \NewM{and} $R_3(\theta,\varphi)$, \NewM{the} $\{g_{00},\,g_{01},\,g_{11}\}$ components of the perturbed metric transform as a scalar, while $\{g_{0i}\}$ and $\{g_{ij}\}$ components (where $i,j=\{2,3\}$) transform as a vector and tensor quantities, respectively\footnote{In four dimensions, after decomposing a symmetric rank-2 tensor into scalars, vectors, and a transverse-traceless tensor, we get that tensor mode vanishes.}. See \cite{DeFelice:2011ka, Motohashi:2011pw, Gannouji:2021oqz} for a more detailed explanation of the Regge-Wheeler formalism.
Based on the spacetime symmetries, we follow a standard decomposition of perturbation variables into a basis of spherical harmonics, $Y_{\ell m}(\theta,\varphi)$, to derive perturbation equations.
Decomposition of the odd-type metric perturbations can be written as
\begin{eqnarray}
     &  & h_{tt}=0,~~~h_{tr}=0,~~~h_{rr}=0,\\
     &  & h_{ta}=\sum_{\ell, m} h^{(0)}_{\ell m}(t,r)E_{ab}\nabla^{b}Y_{\ell m}(\theta,\varphi),    \label{eq:pertV1} \\
     &  & h_{ra}=\sum_{\ell, m} h^{(1)}_{\ell m}(t,r)E_{ab}\nabla^{b}Y_{\ell m}(\theta,\varphi),    \label{eq:pertV2} \\
     &  & h_{ab}=\frac{1}{2}\sum_{\ell, m} h^{(2)}_{\ell m} (t,r)\left[E_{a}^{~c}\nabla_{c}\nabla_{b}Y_{\ell m}(\theta,\varphi)+E_{b}^{~c}\nabla_{c}\nabla_{a}Y_{\ell m}(\theta,\varphi)\right]. 
     \label{odd-ab}
\end{eqnarray}
where $E_{ab}:=\sqrt{\det\gamma}~\epsilon_{ab}$ with $\gamma_{ab}$ being the two-dimensional metric on the sphere and $\epsilon_{ab}$ being the totally antisymmetric symbol with normalization $\epsilon_{\theta \varphi}=1$. 
The expansion coefficients $h^{(0)}_{\ell m}$, $h^{(1)}_{\ell m}$ and $h^{(2)}_{\ell m}$  are not independent. 
We can use the general covariance of GR to vanish some of these functions by using the gauge transformation $x^{\mu}\to x^{\mu}+\xi^{\mu}$, where $\xi^{\mu}$ is an infinitesimal function.
Under this transformation, metric perturbations transform as $h_{\mu\nu} \longrightarrow h'_{\mu\nu} = h_{\mu\nu} - \bar{\nabla}_\mu \xi_\nu - \bar{\nabla}_\nu \xi_\mu$. 
\NewY{The vector $\xi^{\mu}$ can also be decomposed into vector spherical harmonics.}
Restricting to odd-parity modes, it can be written as
 \begin{equation}
\xi_{t}=\xi_{r}=0,~~~\xi_{a}=\sum_{\ell m}\Lambda_{\ell m}(t,r)E_{a}^{~b}\nabla_{b}Y_{\ell m}.
\end{equation}
Thus, the expansion coefficients transform as
\begin{align}
    h_{\ell m}^{(0)} \rightarrow h_{\ell m}^{(0)}-\dot\Lambda_{lm} \,, \qquad
    h_{\ell m}^{(1)} \rightarrow h_{\ell m}^{(1)}-\Lambda_{lm}'+\frac{C'}{C}\Lambda_{lm} \,, \qquad
    h_{\ell m}^{(2)} \rightarrow h_{\ell m}^{(2)}-2\Lambda_{lm} \, .
    \label{eq:odd-transf}
\end{align}

\NewY{
The above relations tell us that we might pick a gauge in which perturbations variables simplify a bit. 
We choose the function $\Lambda_{\ell m}$ such that $h_{\ell m}^{(2)} = 0$. 
This is known as the {\it Regge-Wheeler gauge}. 
However, this choice of $\Lambda_{\ell m}$ is only valid for higher multipoles ($\ell \geq 2$), because the odd-parity perturbations do not have the monopole term $\ell = 0$ (the spherical harmonic $Y_{00}$ is constant), and $h_{\ell m}^{(2)}$ coefficient is vanishing for the dipole term.}

On the other hand, \NewY{since any vector can be decomposed into a divergence part and a divergence-free part,} 
the perturbation for the vector field that contributes to the odd-parity modes is given by
\begin{align}
    \delta A_t = \delta A_r=0 \,, 
    \qquad 
    \delta A_a = \sum_{\ell m} A^{(v)}_{\ell m}(t,r) E_{ab} \nabla^b Y_\ell^m(\theta,\varphi) \, .
    \label{vectorper}
\end{align}

As explained in \cite{Chandrasekhar98}, when studying perturbations of any spherically symmetric system,
without any loss of generality, we can restrict oneself to axisymmetric modes of perturbations if we assume $m = 0$. 
With this simplification, we can rewrite spherical harmonics in terms of Legendre Polynomials.
Also, since modes with different $\ell$ evolve independently, we focus on a specific mode and omit the indices. 

In what follows, we investigate $\ell \geq 2$ and $\ell = 1$ modes separately and discuss the stability of BH solutions. 

\subsection{Second-order action for higher multipoles, \texorpdfstring{$\ell \geq 2$}{l>=2}}

In this section, we will focus in higher multipoles $(\ell \geq 2)$, since dipole mode $\ell = 1$ require
a special treatment. Expanding the action (\ref{action}) to second order in perturbations,
performing integration over the sphere $(\theta, \varphi)$ (see, for example, Appendix B of Ref. \cite{Kase:2018voo}) and multiple integrations by parts,
the resulting second-order action for odd-parity perturbations yields
\begin{equation}
    S_{\rm odd}^{(2)} = \frac{\lambda(2\ell +1)}{4\pi} \int \dd t \, \dd r \; \mathcal{L}_{\rm odd}
    \label{S_odd_sec}
\end{equation}
where the Lagrangian is given by\footnote{\NewY{We have renamed for simplicity of notations $h^{(0)}_{\ell m}\rightarrow h_0$, $h^{(1)}_{\ell m}\rightarrow h_1$, $A^{(v)}_{\ell m}\rightarrow A_v$}}
\begin{equation}
    \begin{aligned}
    \mathcal{L}_{\rm odd} = & 
    {\ca_1} \qty(\dot{\hl}^2 + {\ho'}^2 + \frac{2 C'}{C}{\ho} \dot{\hl} - 2 \ho' \dot{\hl}) 
    + \qty[(\lambda-2) \ca_2 + \frac{(\ca_1 C')'}{C}] \ho^2
    + (\lambda-2) {\ca_3} \hl^2 +
    \\ & 
    (\lambda -2) {\ca_4} \ho \hl
    + \qty[(\lambda-2) \ca_5 + \cb_3 \frac{C'}{C} - \qty(\cb_2\frac{C'}{C})'] \ho \dA
    + (\lambda -2) {\ca_6} \hl \dA +
    \\ & 
    {\cb_1} \dot{\dA} \left(\dot{\hl} - {\ho'} + \frac{C'}{C}{\ho} \right) 
    + {\cb_2} \left( \dot{\hl} \dA' - \ho' \dA' - \frac{C'}{C} \ho' \dA \right) 
    + {\cb_3} {\dA} \left(\dot{\hl} - {\ho'} \right) +
    \\ & 
    (\lambda \cb_4 + \cb_5) \dA^2 + {\cb_6} {\dA'}^2 + {\cb_7} \dot{\dA}^2 + {\cb_8} \dA' \dot{\dA} 
    \, ,
	\end{aligned}
	\label{secLag-1}
\end{equation}
and $\lambda = \ell(\ell+1)$. The coefficients $\ca_i$, $\cb_i$ are functions of $r$ only and their expressions are given in the Appendix \ref{appendix:odd_coeff}. 
For odd-parity modes with $\ell \geq 2$, there are only two dynamical DOFs. 
We can check that in the above Lagrangian the perturbation variable $h_0$ is non-dynamical. 
We can derive a constraint equation for it by varying the action (\ref{S_odd_sec}) with respect to $h_0$. 
However, this is not an algebraic constraint due to the presence of the term ${\ho'}^2$ in the action. 
Using the method of Lagrangian multiplier (commonly used to tread perturbations in scalar-tensor theories), we rewrite the above Lagrangian as
\begin{equation}
    \begin{aligned}
        \mathcal{L}_{\mbox{\scriptsize odd}}  = & 
        {\ca_1} \left\{ 2\vq \left[ \dot{\hl} - \ho' + \frac{C'}{C} \ho + \frac{1}{2\ca_1} \left(\cb_1 \dot{\dA} + {\cb_2} \dA' + \cb_3 \dA\right) \right] - \vq^2\right\} + 
        \\ &
        (\lambda - 2) \left( \ca_2 \ho^2 + \ca_3 \hl^2 + \ca_4 {\ho} {\hl} + \ca_5 {\ho} {\dA} + \ca_6 {\hl} {\dA} \right) -   
        \\ &
        \frac{1}{4\ca_1} \left(\cb_1\dot{\dA} + \cb_2 \dA' + \cb_3 \dA \right)^2 
        + (\lambda \cb_4 + \cb_5) \dA^2 + \cb_6 {\dA'}^2 + \cb_7 \dot{\dA}^2 + \cb_8 \dA' \dot{\dA} \, ,
   \end{aligned}
   \label{secLag-2}
\end{equation}
where $\vq(t,r)$ is an auxiliary field that we have introduced to simplify the Lagrangian.
We can easily check that by substituting the EOM for $\vq(t, r)$ into Eq.(\ref{secLag-2}), 
we recover the original Lagrangian (\ref{S_odd_sec}). 
Now, varying the Lagrangian (\ref{secLag-2}) with respect to $\ho$ and $\hl$ leads to
\begin{align}
    2 {\ca_1} \vq' + \qty(2 \ca_1' + \frac{2 {\ca_1} C'}{C})\vq +(\lambda -2) (2 {\ca_2} {h_0}+{\ca_4} {h_1}+{\ca_5} \dA) = 0 \, ,
    \label{Eq-h0} \\
    (\lambda -2) ({\ca_4} {h_0}+2 {\ca_3} {h_1}+{\ca_6} \dA)-2 {\ca_1} \dot{\vq} = 0 \, .
    \label{Eq-h1}
\end{align}
We solve Eqs. (\ref{Eq-h0}) and (\ref{Eq-h1}) for $\ho$ and $\hl$ to rewrite both perturbation variables in terms of the auxiliary field $\vq(t,r)$. 
Substituting these expressions into the Lagrangian (\ref{secLag-2}) and performing integration by parts, one finds
\begin{equation}
    (\lambda-2){\cal L}_{\rm odd} = 
    K_{ij} \dot{\chi}_i \dot{\chi}_j + 
    L_{ij} \chi'_i \chi'_j + 
    R_{ij} \dot{\chi}_i \chi'_j + 
    T_{ij} \dot{\chi}_i \chi_j + 
    D_{ij} \chi_i' \chi_j +
    M_{ij} \chi_i \chi_j 
    \,,
    \label{eq:Even-Lag-MatrixForm}
\end{equation}
where we have defined the vector $\chi = (\vq, \; \dA)^T$, and 
$\{K_{ij}, \, L_{ij}, \, R_{ij}, \, M_{ij}\}$ are $2\times 2$ symmetric matrices, while $\{ T_{ij} , \, D_{ij}\}$ are antisymmetric. 
This Lagrangian explicitly shows that we only have two DOFs, one associated with gravitational perturbation and another related to vector perturbation.
The non-vanishing components of ${\mathbf K}, {\mathbf L}$, and ${\mathbf R}$ matrices are
\begin{equation}
    \begin{aligned}
    K_{11} & \equiv \alpha_1 = \frac{4 \ca_1^2 \ca_2}{\ca_4^2 - 4 \ca_2 \ca_3} \,, \quad &
    K_{22}  &\equiv (\lambda-2)\alpha_2 = (\lambda -2) \left({\cb_7}-\frac{{\cb_1}^2}{4 {\ca_1}}\right) \,, \\
    L_{11} & \equiv \beta_1 = \frac{\ca_3}{\ca_2} \alpha_1 \,, \quad &
    L_{22}  &\equiv (\lambda-2)\beta_2 = (\lambda -2) \left({\cb_6}-\frac{{\cb_2}^2}{4 {\ca_1}}\right) \,, \\
    R_{11} & \equiv \mu_1 = \frac{\ca_4}{\ca_2}\alpha_1 \,, \quad &
    R_{22}  &\equiv (\lambda -2) \mu_2 =  (\lambda -2) \left({\cb_8}-\frac{{\cb_1} {\cb_2}}{2 {\ca_1}}\right) \,.
	\end{aligned}
	\label{eq:1_set_matrices}
\end{equation}
From these relations, since $\mathbf{K}$ is a diagonal matrix, the no-ghost condition \NewY{(related to the positivity of the kinetic term coefficients of the fields)} reduces to
\begin{equation}
    \alpha_1 > 0 \, \qquad \mbox{and} \qquad \alpha_2 > 0 \,,
    \label{eq:no_ghosts}
\end{equation}
\NewM{where, it is worth noting that $\alpha_2$ depends on $k_i$ and $\tilde{k}_i$ but $\alpha_1$ does not.}
\NewM{On the other hand,} the non-zero components of the other matrices are
\begin{equation}
    \begin{aligned}
    D_{12} & = - D_{21} = \frac{\lambda-2}{2}\qty[
        \frac{2 {\ca_1} ({\ca_4} {\ca_6}-2 {\ca_3} {\ca_5})}{4 {\ca_2} {\ca_3}-{\ca_4}^2}-{\cb_2}]\,,
    \\
    T_{12} & = - T_{21} = \frac{\lambda-2}{2}\qty[
    \frac{2 {\ca_1} ({\ca_4} {\ca_5}-2 {\ca_2} {\ca_6})}{{\ca_4}^2-4 {\ca_2} {\ca_3}}
    -{\cb_1}]\,,
    \\
    M_{11} & = -(\lambda-2) {\ca_1} - \frac{4 \ca_3 \cT (\ca_1 C' + C \ca_1')^2}{C^2 {\ca_1}}
        + \left[\left(\frac{4 {\ca_1} {\ca_3} C'}{C} + 4 {\ca_3} \ca_1'\right) \cT\right]'\,,
   \\
   M_{22} & = (\lambda-2)\qty[
        \lambda \cb_4 +{\cb_5} 
        -\frac{{\cb_3}^2}{4 {\ca_1}} 
        - \frac{\lambda-2}{4\ca_3} \left(\ca_6^2 + \frac{({\ca_4} {\ca_6} -2 {\ca_3} {\ca_5})^2 \cT}{\ca_1}\right) 
   + \qty(\frac{{\cb_2} {\cb_3}}{4 {\ca_1}})'
   ]\,,
   \\
   M_{12} & = M_{21} = \frac{\lambda-2}{2} \qty[
        {\cb_3}-\frac{{\cb_2}'}{2}
        + C^2 \ca_1^2 \qty(\frac{(2 {\ca_3} {\ca_5}-{\ca_4} {\ca_6}) \cT}{C^2 \ca_1^2})'
   ] \, .
	\end{aligned}
	\label{eq:2_set_matrices}
\end{equation}
In order to simplify the last relations, we have defined $\cT \equiv \ca_1/(4 {\ca_2} {\ca_3}-{\ca_4}^2)$.

\subsection{Speed of propagation}
Based on the Lagrangian (\ref{eq:Even-Lag-MatrixForm}), we can infer that odd-parity perturbations do not propagate at the speed of light. 
To compute the speed propagation of vector perturbations, let us move to the Fourier space and consider that the solution to the odd-parity perturbation is of the form $\chi \propto e^{i(\omega t - k r)}$, where $k$ is a comoving wavenumber and $\omega$ is a frequency. 
In the small scale limit, the dispersion relation obtained from (\ref{eq:Even-Lag-MatrixForm}) can be written as 
$$\det( \omega^2 K_{ij} + k^2 L_{ij} - \omega k R_{ij}) = 0 \,.$$
The propagation speed of the gravitational and electromagnetic perturbation along the radial direction in proper time can be derived by substituting\footnote{The propagation speed $c_r$ along the radial direction in proper time outside the horizon is given by $c_r = dr_*/d\tau$, where we have defined $d\tau = \sqrt{A} dt$ and $d r_* = dr/\sqrt{B}$. Since this is related to the propagation speed $\hat{c}_r$ in the coordinates $(t, r)$ as $\hat{c}_r = \sqrt{AB} c_r$, we get $\omega = \sqrt{AB} c_r k$} $\omega = \sqrt{AB} c_r k$ into the dispersion relation. 
Solving for $c_r$, we obtain the propagation of the two DOFs:
\begin{align}
    c_{r1} &= \frac{{\mu_1} \pm \sqrt{{\mu_1}^2-4 {\alpha_1} {\beta_1}}}{2 {\alpha_1} \sqrt{A B}} 
    = \frac{\ca_4 \pm \sqrt{\ca_4^2 - 4 \ca_2 \ca_3}}{2 \ca_2 \sqrt{A B}} \,,
    \label{eq:cr1} \\
    c_{r2} &= \frac{{\mu_2} \pm \sqrt{{\mu_2}^2-4 {\alpha_2} {\beta_2}}}{2 {\alpha_2} \sqrt{A B}}
    = 
    \frac{{\cb_1} {\cb_2}-2 {\ca_1} \left({\cb_8} \pm \sqrt{\mathcal{G}}\right)}{\sqrt{A B} \left( \cb_1^2 - 4 \ca_1 \cb_7 \right)}
    \,,
    \label{eq:cr2} 
\end{align}
\refcom{where we have defined
\begin{equation*}
    \mathcal{G} = \frac{ \cb_2 \qty(\cb_2 \cb_7 - \cb_1 \cb_8) + \cb_6 \left( \cb_1^2 - 4 \ca_1 \cb_7 \right) + \ca_1 \cb_8^2 }{\ca_1}.
\end{equation*}}
Both expressions might be either positive or negative, depending on the direction along which the odd-parity perturbations propagate. To avoid small-scale Laplacian instabilities along the radial direction, we
require the two conditions:
\begin{align}
    \ca_4^2 - 4 {\ca_2} {\ca_3} \geq 0 \, , \qquad
    \mathcal{G} > 0 \,,
    \label{eq:cond_vel}
\end{align}
under which, the first condition in (\ref{eq:no_ghosts}) reduces to $\ca_2 > 0$.
It can be checked that when the vector field is absent, $c_{r2}=0$; therefore, $c_{r1}$ and $c_{r2}$ correspond to the radial speed propagation of the gravity DOF and the vector-field sector, respectively.

\refcom{In the large $\lambda = l(l+1)$ limit, the matrices $\mathbf{M}$ and $\mathbf{T}$ contribute to the propagation speed along the angular direction ($c_\Omega$). And, considering that the solution to the odd-parity perturbation is of the form $\vec{\mathcal{X}}^{t} \propto e^{i(\omega t- \ell \theta)}$, the dispersion relation is given by}
\begin{equation}
{\rm det}(\omega^2\mathbf{K}-i\omega \mathbf{T}+\mathbf{M})=0\,.
\label{det_angularVel}
\end{equation}
The propagation speed along the angular direction in proper time is obtained by substituting the relation
$\omega^2 = A(r) c_{\Omega}^2 \ell^2/C(r)$ into Eq.~(\ref{det_angularVel}) and solve it for $c_{\Omega}^2$. 
Taking the $\ell \to \infty$ limit at the end, 
we obtain the two propagation speed squares as
\begin{equation}
    c_{\Omega_\pm}^2 = 
    \frac{C}{2 A \alpha_1 \alpha_2} \left[
        \ca_1 \alpha_2 - \alpha_1 D_1 + D_2^2 
        \pm 
        \sqrt{(\ca_1 \alpha_2 + \alpha_1 D_1)^2 + D_2^2 (2 \ca_1 \alpha_2 - 2\alpha_1 D_1+D_2^2)}
    \right]
    \label{eq_AngularVel}
\end{equation}
where we have defined
\begin{equation}
    D_1 = {\cb_4}-\frac{{\ca_2} {\ca_6}^2+{\ca_3} {\ca_5}^2-{\ca_4} {\ca_5} {\ca_6}}{4 {\ca_2} {\ca_3}-{\ca_4}^2} 
    \, \qquad
    D_2 = \frac{{\cb_1}}{2}-\frac{{\ca_1} ({\ca_4} {\ca_5}-2 {\ca_2} {\ca_6})}{{\ca_4}^2-4 {\ca_2} {\ca_3}}\, .
\end{equation}
Clearly, to avoid the Laplacian instability for large $\lambda$, we require the conditions
\begin{equation}
    c_{\Omega_\pm}^2 > 0 \, \label{cond_angular}.
\end{equation}

\NewY{Thus, we have shown that the conditions (\ref{eq:no_ghosts}), (\ref{eq:cond_vel}), and (\ref{cond_angular}) need to hold for avoiding ghost and Laplacian instabilities.}

\subsection{The dipole mode \texorpdfstring{$\ell = 1$}{l=1}}

As discussed in the previous sections, the Regge-Weeler gauge is not valid for the dipole mode since perturbation $h_{ab}$ in (\ref{odd-ab}) vanishes.\footnote{\NewY{Fixing $m=0$, since the final result is independent of the azimuthal angle, get that $Y_{10}(\theta, \varphi) \propto \cos(\theta)$; thus, expanding the square brackets in (\ref{odd-ab}), it vanishes automatically.}}
Therefore, we have the freedom to choose another gauge to simplify the calculations. 
\refcom{As we will see, we only have one DOF in this mode associated with the perturbation of the vector field ($A_v$).}
Returning to the transformation of metric perturbations (\ref{eq:odd-transf}), we might use the gauge freedom to fix $h^{(1)}_{\ell m}=0$. But, this choice does not fix the gauge completely. We have a residual gauge freedom defined as $\Lambda\rightarrow\Lambda +f(t)C(r)$. Consequently, \refcom{instead of working at the level of the action,} we fixed the gauge after obtaining the EOMs.

The Lagrangian for this mode is obtained by taking $\ell = 1$ ($\lambda=2$) in Eq. (\ref{secLag-1}). 
This Lagrangian can be written as
\begin{equation}
    \begin{aligned}
        \mathcal{L}_{\rm odd}^{(\ell = 1)} = & 
        {\ca_1} \left[ \dot{\hl} - \ho' + \frac{C'}{C} \ho + \frac{\cb_1 \dot{\dA} + {\cb_2} \dA' + \cb_3 \dA}{2\ca_1}  \right]^2
        - \frac{1}{4\ca_1} \left(\cb_1\dot{\dA} + \cb_2 \dA' + \cb_3 \dA \right)^2 
        \\ &
        + (2 \cb_4 + \cb_5) \dA^2 + \cb_6 {\dA'}^2 + \cb_7 \dot{\dA}^2 + \cb_8 \dA' \dot{\dA}
        \, .
   \end{aligned}
   \label{eq:Lag_l1}
\end{equation}

Variation of the action with respect to $\ho$ and $\hl$ gives us
\begin{equation}
    \partial_t \mathcal{E} = 0 \,, \qquad\qquad \partial_r\qty(C \mathcal{E}) = 0 \,,
\end{equation}
where we have defined\footnote{After fixing the gauge, $h_1=0$.}
\begin{equation}
    \mathcal{E} = 
    \cb_1 \dot{\dA} + \cb_2 \dA' + \cb_3 \dA - 2\ca_1\qty(\ho' - \frac{C'}{C}\ho) \,,
\end{equation}
which solution is given by $\mathcal{E} = \mathcal{J}/C(r)$, where $\mathcal{J}$ is an integration constant related to the angular momentum of a slowly rotating BH.
The above relation can be integrated to obtain
\begin{equation}
    \ho(t,r) = C(r) \int \dd r \frac{1}{2\ca_1C}\qty(\cb_1 \dot{\dA} + \cb_2 \dA' + \cb_3 \dA - \frac{\mathcal{J}}{C} ) + F(t)C(r)
\end{equation}
where $F(t)$ is a constant of integration. This last term can be eliminated with the help of
the residual gauge freedom. Now, the variation of (\ref{eq:Lag_l1}) wrt $\dA$ gives
\begin{equation}
    {\alpha_2} \ddot{\dA} + ({\beta_2} \dA')' + {\mu_2} \dot{\dA}' + \frac12 {\mu_2}' \dot{\dA} 
    - \mathcal{V}(r)\dA
    = \mathcal{S}(r)
    \label{eq:Eq_dA_l1}
\end{equation}
where we have defined
\begin{equation}
    \mathcal{V}(r) = \qty(\frac{{\cb_2} {\cb_3}}{4 {\ca_1}})'-\frac{{\cb_3}^2}{4 {\ca_1}}+2 {\cb_4}+{\cb_5} 
    \, ,
    \qquad
    \mathcal{S}(r) = \frac{\mathcal{J}}{4}\qty[\frac{{\cb_3}}{\ca_1 C} - \qty(\frac{{\cb_2}}{{\ca_1} C})']
    \, .
\end{equation}

\NewY{To avoid ghosts in the dipole mode, we impose the positivity of the kinetic term coefficient, this condition reduces to $\alpha_2 >0$. Therefore, we obtain the same condition as for the higher multipoles.}
From equation (\ref{eq:Eq_dA_l1}), one can show that the radial propagation speed of $\dA(t,r)$ is the same as $c_{r2}$ given by Eq. (\ref{eq:cr2}). Thus, the dipole perturbation does not produce any additional constraint for the absence of ghosts and Laplacian instability to those derived for $\ell \geq 2$. 

\refcom{In the following section, we applied the above results, which are very general, to specific examples encountered in the literature. In particular, our attention focused on the case of $U(1)$ gauge symmetry theory. We extend the results of Ref. \cite{Heisenberg:2018mgr} by computing the stability conditions and deriving wave-like equations from which the QNMs can be obtained.}

\section{Applications to specific theories}

\subsection{SVT theories with \texorpdfstring{$U(1)$}{U(1)} gauge-invariant symmetry}

\NewM{The searching for} a simple condition, which we can determine the stability of BH solutions in the most general case, is not an easy task. As we showed, any relation, including the equations of motion derived from the Lagrangian (\ref{eq:Even-Lag-MatrixForm}), are long expressions and unwieldy.
To simplify the calculation of perturbations, in the following, we review and study particular examples found in the literature.

As a first example, we consider the most general Lagrangian for a scalar field and a $U(1)$ gauge field with derivative interactions.
As shown in \cite{Heisenberg:2018acv}, the most general SVT theory with gauge invariance and second-order equations of motion is given by the action
\begin{equation}
\mathcal{S}_{\rm SVT}= \int \dd^4x \sqrt{-g}\qty[ f_4(\phi) R + \mathcal{L}^{2, U(1)}_{\rm SVT} + \mathcal{L}^{3, U(1)}_{\rm SVT} + \mathcal{L}^{4, U(1)}_{\rm SVT} ] \label{action_U1}
\end{equation}
where\footnote{We keep the same notation for the functions $f_i$ used in the Lagrangian (\ref{eq:Lag_SVT})}
\begin{eqnarray}
\mathcal{L}^{2, U(1)}_{\rm SVT} &=& f_2(\phi, X_1, F) \nonumber \\
\mathcal{L}^{3, U(1)}_{\rm SVT} &=& \left[
    k_1(\phi, X_1)g_{\rho\sigma}  + k_2 (\phi, X_1)\nabla_\rho\phi\nabla_\sigma\phi\right]
    \tilde{F}^{\mu\rho}\tilde{F}^{\nu\sigma} \nabla_\mu\nabla_\nu\phi \label{U1LagrangianSVT} \\
\mathcal{L}^{4, U(1)}_{\rm SVT} &=& 
    f_6(\phi, X_1) L^{\mu\nu\alpha\beta}F_{\mu\nu}F_{\alpha\beta}
    + \frac12 f_{6,X_1} \tilde{F}^{\mu\nu}\tilde{F}^{\alpha\beta} \nabla_\mu\nabla_\alpha\phi\nabla_\nu\nabla_\beta\phi\,.  \nonumber
\end{eqnarray}
Odd-parity perturbations of this theory have been studied in \cite{Heisenberg:2018mgr}. 
Using the above calculations, we get that the matrices $\mathbf{T}$ and $\mathbf{R}$ (\ref{eq:2_set_matrices}) vanishes and the Lagrangian (\ref{eq:Even-Lag-MatrixForm}) simplifies a bit.
\refcom{For the action (\ref{eq:U1_velo}), the stability conditions (\ref{eq:no_ghosts}) reduces to impose 
\begin{eqnarray}
    -B^2 {A_0'}^2 \phi '^2 f_{6, X_1} + B {A_0'}^2 {f_6}-\frac{1}{2} A {f_4} &<& 0 \ ,\\
    C' \phi ' \left(B' \phi '+2 B \phi ''\right) f_{6, X_1} + \frac{k_{1}}{2} \left(\frac{C B' \phi '}{B} + C' \phi '+2 C \phi ''\right)  &&\nonumber \\ 
   \qquad \qquad + \frac{C}{2 B}f_{2, F} + \left(-\frac{B' C'}{B}-2 C'' + \frac{C'^2}{C}\right) f_{6} + \frac{1}{2} B C' \phi '^3 k_{2} &>& 0 \, ,
\end{eqnarray}
which are the same stability conditions derived in the original paper\footnote{Under the appropriate assignment of Lagrangian coefficients.} \cite{Heisenberg:2018mgr}. 
For this particular example, we can go further and determine the stability of this theory.}
The equations of motion for this example are written as
\begin{equation}
    \tilde L_{ij} \chi_j '' + (\tilde L_{ij}' - \tilde D_{ij} ) \chi_j' + (\tilde M_{ij} + \omega^2 \tilde K_{ij} - \frac12 \tilde D_{ij}') \chi_j = 0 \, ,
\end{equation}
where tilde matrices are the matrices (\ref{eq:1_set_matrices}) and (\ref{eq:2_set_matrices}) evaluated with the Lagrangian (\ref{U1LagrangianSVT}). Here, we have used $\vec{\chi}(t,r) \to e^{-i\omega t}\vec{\chi}(r)$. 
Making a change of variable $\chi_i(r) \to S_{ij} (r) \Phi_j(r)$, and changing to tortoise coordinate ($\dd r = \sqrt{AB} \dd r_*$) we get 
\begin{equation}
    \frac{d^2 \vec{\Phi}}{d r_*^2} + \omega^2 \qty( \mathbf{S}^{-1} \mathcal{C}_v^{-1} \mathbf{S})\vec{\Phi} - \mathbf{V} \vec{\Phi} = 0
    \label{eq:U1-EOM-odd}
\end{equation}
where the matrix $ \mathbf{S}$ is solution\footnote{This differential equation will have some integration constants which can be left free as soon as $S$ is invertible.} of 
$
    S'_{ij} + C_{ik}S_{kj}=0 \, ,
$
with the definitions
\begin{align}
    C_{ij} & = \frac12 \tilde L^{-1}_{ik} \qty[\tilde L_{kj}' - \tilde D_{kj} - \frac12 \qty(\frac{A'}{A} + \frac{B'}{B})\tilde L_{kj}] \, , \\
    B_{ij} & = \tilde L_{ik}(C_{km} C_{mj} - C'_{kj}) - \tilde L'_{ik} C_{kj} + \tilde D_{ik} C_{kj} + \tilde M_{ij} - \frac12 \tilde D'_{ij} \, ,
\end{align}
and $\mathcal{C}_v = \mbox{diag}\qty(c_{r1}^2, ~ c_{r2}^2)$, where 
\begin{equation}
    c_{r1}^2 = -\frac{\tilde\ca_3}{AB\tilde\ca_2} = 1 + \frac{2 B^2 f_{6,X_1} {A_0'}^2 \phi'^2}{A f_4 - 2 B {f_6} {A_0'}^2} \,,
    \qquad\quad
    c_{r2}^2 = \frac{\tilde\cb_2^2 - 4 \tilde\ca_1 \tilde\cb_6}{4 AB \tilde\ca_1 \tilde\cb_7} \,,
    \label{eq:U1_velo}
\end{equation}
are the velocities of propagations of the gravitational and vector DOFs. In Eq. (\ref{eq:U1-EOM-odd}),  $\mathbf{V}$ is the matrix potential defined as
\begin{equation}
    \mathbf{V} = - A(r) B(r) \, \mathbf{S}^{-1} \, \tilde{\mathbf{L}}^{-1} \, \mathbf{B} \, \mathbf{S} \, .
\end{equation}

From equation (\ref{eq:U1_velo}), to obtain $c_{r1}^2 = 1$ (gravity DOF traveling at the speed of light), we must impose the condition $f_6 \equiv f_6(\phi)$. 
Generally, the dependence of this function on $X_1$ leads to the values of $c_{r1}^2$ different from 1.

\subsubsection{Restricting to \texorpdfstring{$c_{r1} = c_{r2} = 1$}{c1 = c2 = 1}}

\NewY{Since the detection of gravitational waves by the LIGO-Virgo collaboration, the propagation speed of gravity DOFs has been very constrained \cite{Cornish:2017jml, LIGOScientific:2017zic, Liu:2020slm}. Nevertheless, these constraints rely on the fact that gravitational waves propagate mostly over a FLRW spacetime rather than a spherically symmetric background. 
For the latter case, options to constraint velocities (\ref{eq:cr2}) rely on more theoretical arguments such as Lorentz invariance, unitarity, and analyticity.
Hereafter,} one of the most interesting solutions is the case in which the speed propagation of both DOFs is 1. \NewY{Under this condition, we can see from Eq. (\ref{eq:U1-EOM-odd}) that we recover the standard wave-like equation for perturbations.}

The most simple Lagrangian that gives this solution consist of taking only the $\mathcal{L}_{{\rm SVT}}^{(2)}$ and $\mathcal{L}_{{\rm SVT}}^{(4)}$ with the restriction that $f_4 \equiv f_4(\phi)$. 
Lagrangian $\mathcal{L}_{{\rm SVT}}^{(6)}$, which corresponds to intrinsic vector modes, always modifies the speed of propagation of both DOFs. In contrast,
we preserve the unitary condition if we impose $f_5 = \text{const}$ in the Lagrangian $\mathcal{L}_{{\rm SVT}}^{(5)}$.
Thus, a theory which gives $c_{r1}^2 = c_{r2}^2 = 1$ is given by  
\begin{align}
    & 
    f_4 = f_4(\phi)  \, , \qquad
    f_5 = \beta_0 = \text{const.} \ , \qquad
    f_6 = \tilde{f}_6 = k_i = \tilde{k}_i = 0 \, ,
\end{align}
while the functions $f_2(\phi,X_1,X_2,X_3,F)$, $f_3(\phi, X_3)$ and $\tilde{f}_3(\phi, X_3)$ are arbitrary. 
The second order Lagrangian (\ref{eq:Even-Lag-MatrixForm}) reduces to
\begin{equation}
    (\lambda-2){\cal L}_{\rm odd} = 
    K_{ij} \dot{\chi}_i \dot{\chi}_j + 
    L_{ij} \chi'_i \chi'_j + 
    M_{ij} \chi_i \chi_j 
    \,. 
\end{equation}
The no ghost conditions (\ref{eq:no_ghosts}) gives
\begin{equation}
    K_{11} = \frac{\sqrt{B} C f_4}{2 A^{3/2}} > 0 \, , \qquad
    K_{22} = \frac{f_{2,F}}{2 \sqrt{A B}} >0
    \label{eq:ghost_cond_simple}
\end{equation}
which imposes $f_1(\phi) > 0$ and $f_{2,F} > 0$. This is the same condition of generalized Einstein-Maxwell-scalar theories \cite{Gannouji:2021oqz}.
Performing a change a of variables and changing to tortoise coordinates, we get that the action for this example can be written as 
\begin{equation}
    S \propto \int \dd t \dd r_* \qty[ \qty(\frac{\partial \vec{\Psi}}{\partial t})^2 - \qty(\frac{\partial \vec{\Psi}}{\partial r_*})^2 - \mathbf{V}(r)\vec{\Psi}^2] \,,
    \label{S2-action}
\end{equation}
where $\vec{\Psi}$ is the new variable that comprises the two available DOFs and $\mathbf{V}(r)$ is the is matrix potential, where
\begin{align}
    V_{11} & = (\lambda -2)\frac{A}{C} - \dv{S_1}{r_*} + S_1^2 \,, \qquad  \qquad \qquad
    S_1(r) = \frac{1}{2} \sqrt{A B} \left(\frac{C'}{C}+\frac{f_4'}{f_4}\right) \\
    V_{22} & = \frac{B{A_0'}^2}{f_4} f_{2,F} + \tilde{V}_2 - \dv{S_2}{r_*} + S_2^2 \,, \,\qquad\quad
    S_2(r) = -\frac{1}{2} \sqrt{A B} \frac{(f_{2,F})'}{f_{2,F}} \\
    V_{12} & = V_{21} = \sqrt{\frac{(\lambda -2) A B A_0'^2 f_{2,F}}{C f_4}}
\end{align}
where we have defined
\begin{align}
    \tilde{V}_2 & = 
    \lambda\frac{A}{C} +
    \frac{A}{f_{2,F}} \left[ f_{2,X_3} +
    2 B A_1 \tilde{f}_3'  +
    \frac{(A B C^2 A_1^2)'}{A C^2 A_1}(f_{3,X_3} + \tilde{f}_3) \, .
    \right] \label{V2_tilde}
\end{align}
\NewY{Again, a prime indicates derivative with respect to the radial coordinate.}
The introduction of the functions $(S_1,S_2)$ will be more transparent in the stability analysis section. 
Variation of the action (\ref{S2-action}) with respect to $\Psi$ gives us a wave-like equation
\begin{equation}
    -\frac{\partial^2 \vec{\Psi}}{\partial t^2} + \frac{\partial^2 \vec{\Psi}}{\partial r_*^2} - \mathbf{V}\vec{\Psi} = 0 \, .
    \label{System-Eq}
\end{equation}
We can see that both modes propagate at the speed of light.
Since matrix $\mathbf{V}$ is not diagonal, the previous equation is a set of two coupled differential equations.

\subsubsection{Stability analysis}

Due to the complex form of the second-order Lagrangian (\ref{eq:Even-Lag-MatrixForm}) for a general SVT theory, we restrict the stability analysis for the action (\ref{S2-action}).
It is well-known that a BH reacts to an external perturbation by emitting a signal in the form of waves, the gravitational radiation.
Performing a Fourier transform of our variables $(\vec{\Psi}\rightarrow e^{-i\omega t} \vec{\Psi})$, we get that 
Eq. (\ref{System-Eq}) is recast as
\begin{align}
    \mathcal{H}\vec{\Psi}=\omega^2\vec{\Psi}
    \label{eq:QNM_eq}
\end{align}
where $\mathcal{H}=-\partial_{r_*}^2+\mathbf{V}$. 
The frequency $\omega^2$ appears as the eigenvalues of the operator $\mathcal{H}$.
After imposing physical boundary conditions to Eq. (\ref{eq:QNM_eq}), the frequencies form a set of discrete complex quantities, called quasi-normal modes (QNMs).

Stability of the solutions means that no perturbation grows unbounded in time. 
In terms of the frequency, unstable modes are equivalent to purely imaginary modes $\omega^2<0$ (see e.g. \cite{Ganguly18}), therefore the stability of the spacetime is related to the positivity of the operator $\mathcal{H}$, namely that $\mathcal{H}$ has no negative spectra. 
As showed in \cite{Gannouji:2021oqz}, to prove the stability, let us define the inner product 
\begin{align}
    (\vec{\psi},\vec{\xi})=\int {\rm d}r_* \Bigl[\bar{\psi}_1 \xi_1 + \bar{\psi}_2 \xi_2 \Bigr] \, ,
    \label{eq:inner_product}
\end{align}
where $\vec{\psi}=(\psi_1,\psi_2)^T$ and $\vec{\xi}=(\xi_1,\xi_2)^T$. 
Stability means that the operator $\mathcal H$ is a positive self-adjoint operator in $L^2(r_*)$ ---the Hilbert space of square integrable functions of $r_*$. 
Therefore, we need to prove the positivity defined as
\begin{align}
    \forall \chi\,,\quad (\vec{\chi},\mathcal{H}\vec{\chi})>0 \, .
    \label{eq:Stability_def}
\end{align}
This condition will imply that given well-behaved initial data, of compact support, $\chi$ remains bounded for all time. 
This is a sufficient condition. 
The rigorous and complete proof of the stability related to equations of the form $(\ref{System-Eq})$ can be found in \cite{wald1,wald2} using spectral theory.

 Using (\ref{eq:inner_product}), we have
\begin{align*}
    (\vec{\chi},\mathcal{H}\vec{\chi}) &=\int {\rm d}r_*\Bigl[\bar{\chi}_1\Bigl(-\partial_{r_*}^2\chi_1+V_{11}\chi_1+V_{12}\chi_2\Bigr)+\bar{\chi}_2\Bigl(-\partial_{r_*}^2\chi_2+V_{22}\chi_2+V_{12}\chi_1\Bigr)\Bigr]
    \\
    &=\int {\rm d}r_*\Bigl[\Big|\frac{{\rm d}\chi_1}{{\rm d}r_*}\Bigr|^2+\Big|\frac{{\rm d}\chi_2}{{\rm d}r_*}\Bigr|^2+V_{12}\Bigl(\bar{\chi}_1\chi_2+\chi_1\bar{\chi}_2\Bigr)+V_{11}|\chi_1|^2+V_{22}|\chi_2|^2\Bigr]
    \\
    & = \int {\rm d}r_*\Bigl[\Big|\frac{{\rm d}\chi_1}{{\rm d}r_*}+S_1\chi_1\Bigr|^2+\Big|\frac{{\rm d}\chi_2}{{\rm d}r_*}+S_2\chi_2\Bigr|^2+(\lambda-2)\frac{A}{C}|\chi_1|^2+\tilde{V}_2|\chi_2|^2 \\ 
    & \qquad +V_{12}\Bigl(\bar{\chi}_1\chi_2+\chi_1\bar{\chi}_2\Bigr)\Bigr]\\
    & = \int {\rm d}r_*\Bigl[\Big|\frac{{\rm d}\chi_1}{{\rm d}r_*}+S_1\chi_1\Bigr|^2+\Big|\frac{{\rm d}\chi_2}{{\rm d}r_*}+S_2\chi_2\Bigr|^2+ \Big|\sqrt{(\lambda-2)\frac{A}{C}} \chi_1 + \sqrt{\frac{B A_0'^2 f_{2,F}}{f_4}} \chi_2\Big|^2 
    \\ & \qquad 
    + \tilde{V}_2|\chi_2|^2 \Bigr]
\end{align*}
where $\bar{\chi}_i$ ($i=1,2$) represent the complex conjugate of $\chi_i$ and should not be confused with background quantities.
In the second line, we have neglected the boundary term $\bar{\chi}_1\partial_{r_*}\chi_1+\bar{\chi}_2\partial_{r_*}\chi_2$ coming from the integration by parts, because we assumed $\chi_1$ and $\chi_2$ to be smooth functions of compact support, while in the third line we have neglected the boundary term $S_1|\chi_1|^2+S_2|\chi_2|^2$. 
\refcom{
In order to obtain $(\vec{\chi},\mathcal{H}\vec{\chi}) \geq 0$ we require $\tilde{V}_2 > 0$. 
Using the no-ghost conditions (\ref{eq:ghost_cond_simple}), we get that $V_{11} > 0$ and $V_{12} > 0$.
Thus, the above condition reduces to show the positivity of the Potential matrix.}

\subsection{GR and generalized Einstein-Maxwell-scalar theories}

As a first example, we consider GR with an electromagnetic field to obtain electrically charged black holes. 
The functions $f_i$ in the action (\ref{action}) are fixed to
\begin{align}
    f_2 = F             \, , \qquad
    f_4 = \frac12       \, , \qquad
    f_3 = \tilde{f}_3 = f_5 = \tilde{f}_5 = f_6 = \tilde{f}_6 = k_i = \tilde{k}_i = 0 \, .
\end{align}
Solving background equations, with boundary conditions $A = B = 1$ at $r \to \infty$, we obtain the Reissner-Nordström (RN) solution
\begin{equation}
    A(r)   = B(r) = 1 - \frac{2M}{r} + \frac{Q^2}{2 r^2} \, , \qquad\quad 
    A_0(r) = P + \frac{Q}{r} \, ,
\end{equation}
where $M$ and $Q$ are the mass and electric charge, respectively, and $P$ is an arbitrary constant that we set to zero in the following calculations.
Clearly, the RN solution suffers neither ghost nor Laplacian instabilities, we can see this directly from conditions (\ref{eq:no_ghosts})
\begin{equation}
    \alpha_1 = \frac{r^2}{2}  A(r) > 0 \,, \qquad 
    \alpha_2 = \frac12 A(r) > 0 \,.
\end{equation}
\refcom{Clearly, these conditions are satisfied only outside the event horizon, where $A(r) > 0$.}
Also, from (\ref{eq:cr2}), we get that all DOFs propagates at the speed of light $c_{r1}^2 = c_{r2}^2 = 1$.
\New{Also, from (\ref{eq_AngularVel}), we get that propagation speeds along the angular direction are $c_{\Omega_\pm}^2 = 1$. 
Since all velocities of propagation are equal to the speed of light, relation (\ref{S2-action}) is valid, and the potential is given by
\begin{equation}
    V(r) = \frac{A(r)}{r^4} \left(
    \begin{array}{cc}
         r (\lambda  r-6 M)+2 Q^2 & -\sqrt{2(\lambda -2)} Q r \\
        -\sqrt{2(\lambda -2)} Q r & 2 Q^2+\lambda  r^2 \\
    \end{array}
    \right) \, .
\end{equation}
It can be checked that this potential is positively defined, thus relation (\ref{eq:Stability_def}) is satisfied and the RN solution is classical stable.
In fact, diagonalizing the above potential (since the eigenvectors are independent of $r$) we get the potential derived originally in \cite{Moncrief74-a}, using this expression and computing the QNMs we get that $\omega_I < 0$ for all values of $l$. 
}

Going a bit further, taking a more general action, where the functions are 
\begin{equation}
    \begin{aligned}
    & f_2 = f_2 (\phi, X_1, F) \, , \qquad
    f_4 = f_1(\phi) \\
    & f_3 = \tilde{f}_3 = f_5 = \tilde{f}_5 = f_6 = \tilde{f}_6 = k_i = \tilde{k}_i = 0 \, .
\end{aligned}
\label{eq:EMS_theory}
\end{equation}
we get the Einstein-Maxwell-scalar solutions. 
With the above calculations, we recover the same results shown in the odd-parity perturbation section of Ref. \cite{Gannouji:2021oqz}.
Specifically, the no ghost condition (\ref{eq:no_ghosts}) reads
\begin{equation}
    \alpha_1 = \frac{\sqrt{B} C f_1}{4 A^{3/2}} > 0 \,, \qquad \quad
    \alpha_2 = \frac{\partial_F f_2 (\phi, X_1, F)}{2 \sqrt{AB}} > 0 \,,
\end{equation}
which imposes $f_1(\phi) > 0$ and $\partial_F f_2(\phi, X, F) > 0$. 
Under these conditions, equations (\ref{eq:cond_vel}) are easily satisfy, and all velocities reduces to $c_{r1}^2 = c_{r2}^2 = c_{\Omega_\pm}^2 = 1$.
\New{Using the above results and relations (\ref{eq:EMS_theory}), we get $\tilde{V}_2 = \lambda A/C > 0$ thus, all Einstein-Maxwell-scalar solutions are classical stable when studying odd-parity perturbations.}

\section{Conclusions}

In this paper, we have studied the stability of spherically symmetric black holes under odd-parity perturbations in the most general SVT theory by expanding the action to the second order in linear perturbations. 
After deriving the master equation for the odd-parity perturbations, we found the conditions to avoid ghost or Laplacian instabilities, Eq. \eqref{eq:no_ghosts}, $\alpha_1 > 0$ and $\alpha_2 > 0$. 
As expected, we obtain that propagating DOFs travel with velocities different from the speed of light. 
For the most general case, we compute the propagation speeds in the radial direction, $c_{r1}$ and $c_{r2}$, and the angular direction $c^2_{\Omega\pm}$, which imposes conditions for higher modes, large $\lambda$. 
Therefore, conditions \eqref{eq:cond_vel} and \eqref{cond_angular} must be satisfied in order to ensure the correct propagation of gravity and the vector field on the spacetime.

\NewM{Once we found the conditions for} the most general case, we have 
\NewM{utilized this formalism to particular theories}.
In our first case, we studied the more general SVT theory with $U(1)$ gauge invariance, Eq. \eqref{action_U1}. Since the velocities of the available DOF are not equal to the speed of light, analyzing the equation of motion in the Fourier space, we obtained that perturbations satisfy a similar wake-like equation (\ref{eq:U1-EOM-odd}), but the coefficient attached to the frequency term depends on the radial coordinate. 
Since a similar action has been studied in \cite{Heisenberg:2018mgr}, we compared our results with the previous one found in the literature. 

Afterwards, we studied the particular case of the speed propagation of both DOFs equal to 1, $c_{r1} = c_{r2} = 1$. This unitary condition is preserved if we impose $\mathcal{L}_{{\rm SVT}}^{(6)}=0$, $f_5 = \text{const}$ and $k_i = \tilde{k}_i = 0$ in $\mathcal{L}_{{\rm SVT}}^{(5)}$, and $f_4 = f_4(\phi)$ in $\mathcal{L}_{{\rm SVT}}^{(4)}$, while the functions $f_2$, $f_3$ and $\tilde{f}_3$ are arbitrary. For this case, we found the same no-ghost condition of generalized Einstein-Maxwell-scalar theories, $f_1(\phi) > 0$ and $f_{2,F} > 0$. Even more, we use a generalized Regge-Wheeler potential for a stability analysis of quasi-normal modes, finding the stability condition $\tilde{V}_2>0$, where $\tilde{V}_2$ is given by equation \eqref{V2_tilde}.

Finally, we have applied our results to better known theories such as GR and generalized Einstein-Maxwell-scalar theories. Both theories have DOFs that travel at the speed of light and are stable under odd-parity perturbations.

\bigskip

\textbf{Acknowledgments:} The authors would like to thank Radouane Gannouji for useful discussions and for comments on the manuscript. Yolbeiker Rodr\'{\i}guez Baez and M. Gonzalez-Espinoza acknowledge VRIEA-PUCV for financial support through Proyecto Postdoctorado 2022 VRIEA-PUCV.

\newpage
\appendix 

\section{Odd-parity Lagrangian coefficients \label{appendix:odd_coeff}}

The coefficients which appear in the second-order Lagrangians (\ref{secLag-1}) and (\ref{secLag-2}) are
\begin{align*}
a_1 & = \frac{\sqrt{B}}{4 C\sqrt{A}} \qty[
    2C {f_4} - 4 C X_3 f_{4, X_3} + C' B A_1 X_3 f_{5,X_3} - A_1 B C \phi' f_{5,\phi}]
\\
a_2 & = \frac{\sqrt{B}}{C\sqrt{A}} \biggl[
    \frac{{f_4}}{2 B}
    - \frac{A_0^2 A_0'^2}{A^2} \tilde{f}_{6, X_3} 
    + \frac{1}{4} \left[\frac{A_0}{A \sqrt{B}}\left(A_0 A_1 \sqrt{B}\right)' - \frac{1}{2} A_1 \left(A_1^2 B\right) '\right] f_{5,X_3}
    \\ & 
    - \frac{A_0^2}{2 A B} f_{4, X_3}
    + \frac{1}{4} A_1 \phi ' f_{5,\phi} 
    - \frac{A_0'^2}{A}(f_ 6 + \tilde{f}_6)
    \bigg]
\\
a_3 & = \frac{\sqrt{B}}{C\sqrt{A}} \biggl[
    - \frac{A}{2} {f_4}
    + \frac{A_1 B}{4}\left(A_0 A_0' - X_3 A'\right) f_{5,X_3}
    + \frac{A B A_1}{4} \qty(\phi' f_{5,\phi}  - 2 A_1 f_{4, X_3})
    \\ & \quad
    - B^2 A_0'^2 \qty(A_1^2 \tilde{f}_{6, X_3} + \phi'^2 f_{6, X_1})
    + B A_0'^2 ({f_6} + \tilde{f}_6)
   \bigg]
\\
a_4 & = \frac{\sqrt{B}}{C\sqrt{A}} A_0 \bigg[
    \frac{1}{2 A} \left(X_3 A' - A_0 A_0' \right) f_{5,X_3}
    + \frac{2 A_1 B A_0'^2}{A} \tilde{f}_{6, X_3}
    + A_1 f_{4, X_3}
    - \frac{\phi'}{2} f_{5,\phi}
    \bigg]
\\
a_5 & = \frac{\sqrt{B}}{C\sqrt{A}} \biggl\{
    -\frac{A_0 }{B} f_{4, X_3}
    + A_0' \qty[A \qty(\frac{A_0^2}{A^2})' - 2(A_1^2 B)' ] \tilde{f}_{6, X_3}
    + 2 A_0' ({f_6} + \tilde{f}_6) \left(\frac{B'}{B} - \frac{A'}{A} + \frac{2 A_0''}{A_0'}\right) 
    \\ & \quad
    + \frac{A_0 A_1}{4} \left(\frac{A'}{A} + \frac{B'}{B} + \frac{2 A_1'}{A_1}\right) f_{5,X_3}
    - 2 f_{6, X_1} A_0' \phi ' \left(B' \phi '+2 B \phi ''\right) 
    + 4 A_0' \phi ' \left(\tilde{f}_{6, \phi}+f_{6, \phi}\right)
    \bigg\}
\\
a_6 & = \frac{\sqrt{B}}{C\sqrt{A}} \bigg[
    \frac12 \qty(X_3 A' - A_0 A_0') f_{5,X_3} + A A_1 f_{4, X_3} - \frac{\phi'}{2} A f_{5,\phi} + 2 B A_1 A_0'^2 \tilde{f}_{6, X_3} \bigg]
\\
b_1 & = \frac{\sqrt{B}}{C\sqrt{A}} \biggl[
    A_1 B C' \qty(\frac{A_1}{4} f_{5, X_3}
    -\frac{2 A_0 A_0'}{A} \tilde{f}_{6, X_3})
    - A_1 C f_{4, X_3} 
    + \frac{\phi'}{2} C f_{5, \phi}
    - \frac{2 A_0 C A_0'}{A} \tilde{k}_1
    \biggl]
\\
b_2 & = \frac{\sqrt{B}}{C\sqrt{A}} \biggl[
    A_0 C f_{4, X_3}
    + 2 B^2 A_0' C' \qty(A_1^2  \tilde{f}_{6, X_3} + \phi '^2 f_{6, X_1} )
    -\frac{1}{4} A_0 A_1 B C' f_{5, X_3}
    \\ & \qquad
    + B C A_0'\qty(2 A_1 \tilde{k}_1 + \phi ' k_1)
    -2 B  A_0' C' ({f_6} + \tilde{f}_6)
    \biggl]
\\
b_3 & = \frac{\sqrt{B}}{C\sqrt{A}} \biggl[
    - \frac{2 A_1^2 B^2 A_0' C'^2}{C} \tilde{f}_{6, X_3}
    + \frac{A_1 B}{4}\left(\frac{A_0 C'^2}{C} - A_0' C'\right)f_{5, X_3}
    - \frac{2 B^2  A_0' C'^2 \phi'^2}{C} f_{6, X_1}
    \\ & \qquad
    + f_{4, X_3}\left(C A_0'- A_0 C'\right)
    - C f_{2, F} A_0'
    + \frac{2 B A_0' C'^2}{C} ( f_6 + \tilde{f}_{6})
    - 2 A_0' C' B \qty(2 A_1 \tilde{k}_1 + \phi' k_1 )
    \biggl]
\\
b_4 & = \frac{\sqrt{B}}{C\sqrt{A}} \biggl\{
    - \frac{A}{2 B} f_{2, F} 
    - A' \phi ' \left(B' \phi ' + 2 B \phi ''\right) f_{6, X_1}
    + \left[A' \left(\frac{A_0^2}{A}- A_1^2 B\right)' - 2 A_0'^2 \right] \tilde{f}_{6, X_3}
    \\ & \quad
    + \frac{1}{2} \left[A_0^2 \left(\frac{B' \phi '}{B} + 2 \phi ''\right) - A_1^2 B A' \phi'\right] k_3
    + \left[\frac{A_0^4}{A_1 A^2 B} \qty(\frac{A_1^2 A^2 B}{A_0^2})' - A_1^3 B A'\right] \tilde{k}_3
    \\ & \quad
    + \phi'\left(2 X_3 A' - A_0 A_0'\right) \tilde{k}_4
    - k_1 \left(\frac{(AB)'\phi '}{2 B} + A \phi''\right)
    - \left(\frac{A_1 (AB)'}{B} + 2 A A_1'\right) \tilde{k}_1
    \\ & \quad
    - \frac12 B A' \phi'^2 \qty(2 A_1 \tilde{k}_2 + A_1 k_4 + \phi' k_2)
    + (f_6 + \tilde{f}_{6}) \left(\frac{A' B'}{B} - \frac{A'^2}{A} + 2 A''\right)
    \biggl\}
\end{align*}
\newpage
\begin{align*}
b_5 & = \frac{\sqrt{B}}{C\sqrt{A}} \biggl\{
    \frac{A'C}{2} f_{4, X_3}'
    + \left[\sqrt{AC}\left(\sqrt{AC}\right)'' + \frac{B'}{4B}(A C)' - \frac{A' C'}{4}\right] f_{4, X_3}
    - \frac{(AC)'\phi'}{2} f_{4, \phi X_3}
    \\ &
    - \frac{A C'}{4} \qty[ \frac{\qty(A_1^2 B C)'}{C} + \qty(\frac{A_0^2}{A})' + 2 A_1^2 B \frac{A'}{A}] f_{4, X_3 X_3}
    + \left[\frac{A_1 A' C' B^2}{16 A} \qty(\frac{A}{B})' - \frac{\left(A_1 B A' C'\right)'}{8}\right] f_{5, X_3}
    \\ &
    + \frac{A_1 B C' \phi ' (AC)'}{8 C} f_{5, \phi X_3}
    + \frac{A_1 B C'}{8 C} \left[\frac{A'}{2} \qty(B A_1^2 C - \frac{A_0^2 C}{A})' + A_0 A_0' C' \right] f_{5, X_3 X_3}
    \\ &
    - \qty[\frac{A_1}{2 B}\left(B \left(C A'+2 A C'\right)+A C B'\right) + A C A_1' ]f_{3, X_3}
    - A A_1 C \phi' \tilde{f}_{3, \phi}
    - \frac{A C}{2 B} f_{2, X_3}
    \\ &
    + \frac{B A_0'^2 C'^2}{2 C} \tilde{f}_{6, X_3} 
    - \frac{(A_1 B A_0'C')^2}{2 C} \tilde{f}_{6, X_3 X_3}
    - \sqrt{\frac{A}{B}} \left(\sqrt{A B} C A_1\right)' \tilde{f}_3
    \\ &
    -\frac{1}{2} B A_0'^2 C' \left(2 A_1 \tilde{k}_{1, X_3} + \phi' k_{1, X_3} \right)
    + \frac{A_1 A_0' C'}{2 C} \left(2 B \left(C A_0'-A_0 C'\right)+A_0 C B'\right) \tilde{k}_3
    \\ &
    + \sqrt{A} \qty[\frac{A_0 B A_0' C' \phi'}{2\sqrt{A}} k_3 + \frac{A_0 A_1 B A_0' C'}{\sqrt{A}} \tilde{k}_3 ]'
    + \frac{A_0' C' \phi ' C^2}{4 A_0} \qty(\frac{A_0^2 B}{C^2} {k_3})'
    \biggl\}
\\
b_6 & = \frac{\sqrt{B}}{C\sqrt{A}} \biggl[
    - A' C' B^2 \qty(A_1^2 \tilde{f}_{6, X_3} + \phi'^2 f_{6, X_1} )
    - \frac{1}{2} A C f_{2, F}
    - A_1 B \tilde{k}_1 \left(C A'
    + A C'\right)
    \\ & \quad 
    + B A' C' (f_6 + \tilde{f}_{6})
    - \frac{1}{2} B k_1 \phi ' \left(C A'
    + A C'\right)
    + \frac{1}{2} A_0^2 B C' (2 A_1 \tilde{k}_3 + \phi ' k_3 )
    \biggl]
\\
b_7 & = \frac{\sqrt{B}}{C\sqrt{A}} \biggl\{
    \frac{C}{2 B} f_{2, F} 
    + A_1 C' \left(2 B A_1' + A_1 B'\right) \tilde{f}_{6, X_3}
    + C' \phi ' \left(B' \phi ' + 2 B \phi ''\right) f_{6, X_1}
    + \frac{B}{2} C' \phi'^3 k_2
    \\ & \quad
    + \tilde{k}_1 \left[C \left(2 A_1' + \frac{A_1 B'}{B}\right) + A_1 C'\right]
    + A_1^3 B \tilde{k}_3 C'
    + (f_6 + \tilde{f}_{6}) \left(\frac{C'^2}{C} - \frac{B' C'}{B} - 2 C''\right)
    \\ & \quad
    + \frac{1}{2} A_1 B C' \phi' \qty(2 \phi' \tilde{k}_2 + A_1 k_3 + \phi' k_4 + 2 A_1 \tilde{k}_4)
    + k_1 \left[\frac{\phi '}{2 B} \left(C B' + B C'\right) + C \phi ''\right]
    \biggl\}
\\
b_8 & = \frac{\sqrt{B}}{C\sqrt{A}} \biggl[
    \frac{2}{A}\left(A_0 A' - A A_0'\right) \qty(A_1 B C' \tilde{f}_{6, X_3} +  C \tilde{k}_1)
    - B C' A_0 A_1 \qty(2  A_1  \tilde{k}_3 + \phi' \tilde{k}_4 + \phi' k_3)
    \\ & \quad 
    - \frac{1}{2} A_0 B k_4 C' \phi'^2
    \biggl]
\end{align*}
\normalsize

\newpage

\providecommand{\href}[2]{#2}\begingroup\raggedright\endgroup

\end{document}